\documentstyle[12pt]{article}
\newlength{\abstractwidth}
\begin{document}
\thispagestyle{empty}
\pagestyle{plain}
\renewcommand{\thefootnote}{\fnsymbol{footnote}}
\renewcommand{\thanks}[1]{\footnote{#1}} 
\newcommand{\starttext}{
\setcounter{footnote}{0}
\renewcommand{\thefootnote}{\arabic{footnote}}}

\begin{titlepage}
\bigskip
\hskip 3.7in\vbox{\baselineskip12pt
\hbox{NSF-ITP-95-122}\hbox{hep-th/9510017}}
\bigskip\bigskip\bigskip\bigskip

\centerline{\large \bf Dirichlet-Branes and Ramond-Ramond Charges}

\bigskip\bigskip
\bigskip\bigskip

\centerline{\bf Joseph Polchinski\thanks{joep@itp.ucsb.edu}}
\medskip
\centerline{Institute for Theoretical Physics}
\centerline{University of California}
\centerline{Santa Barbara, CA\ \ 93106-4030}

\bigskip\bigskip

\begin{abstract}
\baselineskip=16pt
We show that Dirichlet-branes, extended objects defined by mixed
Dirichlet-Neumann boundary conditions in string theory, break half of the
supersymmetries of the type~II superstring and carry a complete set of
electric and magnetic Ramond-Ramond charges.  We also find that the product of
the electric and magnetic charges is a single Dirac unit, and that the quantum
of charge takes the value required by string duality.  This is strong evidence
that the Dirchlet-branes are intrinsic to type II string theory and are the
Ramond-Ramond sources required by string duality.  We also note the existence
of a previously overlooked 9-form potential in the IIa string, which gives
rise to an effective cosmological constant of undetermined magnitude.
\end{abstract}
\end{titlepage}
\starttext
\baselineskip=18pt
\setcounter{footnote}{0}

The type~II closed superstring has two kinds of gauge field, from the
Neveu-Schwarz--Neveu-Schwarz (NS\,NS) and Ramond--Ramond (RR) sectors of the
string Hilbert space.\footnote{For a review
of string theory see ref.~\cite{GSW}.}
The respective vertex operators are
\begin{eqnarray}
&& j \bar\partial X^\mu A_\mu(X)\\
&& \bar Q \Gamma^{[\mu_1} \ldots \Gamma^{\mu_n]} Q F_{\mu_1 \ldots \mu_n}(X).
\end{eqnarray}
Here $j$ is a world-sheet weight $(1,0)$ current and $\bar Q_\alpha$ and
$Q_\alpha$ are $(0,1)$ and $(1,0)$ spin fields, the world-sheet currents
associated with spacetime supersymmetry~\cite{FMS}.  From the physical state
conditions,
$A_\mu(X)$ plays the role of a spacetime vector potential, while the physical
state conditions for $F$ imply (in the notation of forms)
\begin{equation}
dF = d^* F = 0.
\end{equation}
These are the Bianchi identity and field equation for an $n$-form field
strength.\footnote
{As an aside, if one considers the Ramond generators in a linear dilaton
background, one sees that the Bianchi identity and field equation contain a
term proportional to the dilaton gradient.  In order to obtain the standard
equations one must rescale by an exponential of the dilaton.  The spacetime
action for the field $F$ appearing in the vertex operator is multiplied by
the usual $e^{-2\phi}$, whereas the field after rescaling has a
dilaton-independent action.  This is the world-sheet explanation of the latter
much-noted fact.}

The NS\,NS and RR gauge fields are quite different in perturbation theory.
String states carry the world-sheet charge associated with the current $j$,
and this translates into a charge under the corresponding NS\,NS spacetime
gauge symmetry.  On the other hand, all string states are neutral under the
RR symmetries because only the field strength $F$ appears in the vertex
operator.  Further, backgrounds with nontrivial NS\,NS gauge fields are
well-studied in conformal field theory, whereas backgrounds of RR gauge
fields are not easily understood in this way: the spin fields depend on the
ghosts, with the additional complication of picture-changing, and they break
the separate superconformal invariances of the matter and ghost theories.

One of the important lessons of string duality is that such world-sheet
distinctions are artifacts of string perturbation theory, with no
invariant significance.  Various dualities interchange NS\,NS and RR states,
and string duality requires that states carrying the various RR charges
exist~\cite{RRch}.  Previously it has been suggested that these are
black $p$-branes, extended versions of black
holes~\cite{black}.  In this paper we will observe that there is another class
of objects which carry the RR charges, the D(irichlet)-branes studied in
ref.~\cite{DLP}.

Let us begin with a type~II closed superstring theory.  Add open strings
with Neumann boundary conditions on $p+1$ coordinates and Dirichlet
conditions on the remaining $9-p$,
\begin{eqnarray}
n^a \partial_a X^\mu &=& 0, \quad \mu = 0, \ldots, p \nonumber\\
X^\mu &=& 0, \quad \mu = p+1, \ldots, 9.  \label{mixed}
\end{eqnarray}
The open string endpoints thus live on a hyperplane, the D-brane, with $p$
spatial and one timelike dimension.  Only closed strings propagate in the
bulk of spacetime,  but sense the hyperplane through the usual open-closed
interactions. This is a consistent string theory, provided $p$ is even in the
IIa theory or odd in the IIb theory.  The consistency conditions will be
explained further below, but consistency can also be seen from the fact that
these boundary conditions arise in the $T$-dual of the usual type~I string
theory~\cite{DLP,odual}.

One would not expect a perfectly rigid object in a theory with gravity, and
indeed the D-brane is dynamical.  In ref.~\cite{DLP} it is shown that there
are massless open-string excitations propagating on the D-brane, the
$T$-duals of the photons, with precisely the properties of collective
coordinates for transverse fluctuations of the D-brane.  It is further
shown that since the D-brane tension arises from the disk, it scales
in string units as $g^{-1}$, $g$ being the closed string coupling.
This is the same coupling-constant dependence as for the branes carrying RR
charges.\footnote{After Edward Witten's talk at Strings '95, Michael Green and
the author both noted this parallel, but various mental blocks prevented the
next step.  Some of the present work is anticipated in
refs.~\cite{gdinst,pdinst}.}  Now let us take this further.  Far from the
D-brane we see only the closed-string spectrum, with two $d=10$ gravitinos.
However, world-sheet boundaries reflect the right-moving
$Q_\alpha$ into the left-moving $\bar Q_\alpha$, so only one linear
combination of the two supercharges is a good symmetry of the full state.
In other words, in the type II theory coupled to the D-brane, half of the
supersymmetries of the bulk theory are broken: {this is a BPS state}.

The BPS property and the scaling of the tension identify the D-brane as a
carrier of RR charge, but we can also see this by direction calculation.
The disk tadpole for a closed string state $|\psi\rangle$ can be written as
$\langle\psi|B\rangle$ where $|B\rangle$ is the closed-string state
created by the boundary~\cite{callan,polcai,horava}.\footnote
{The reader need not feel compelled to work through these rather detailed
references: the essential points are evident in the simple
calculation~(\ref{openloop}).}
In ref.~\cite{polcai,horava} this is studied for the RR sector of
the superstring with Neumann boundaries, and in ref.~\cite{gdinst} for fully
Dirichlet conditions.  The Ramond ground-state component of $|B\rangle$ is
determined by a condition
\begin{equation}
(\psi^\mu_0 - \tilde\psi^\mu_{0}) |B\rangle = 0,\quad \mu=0,\ldots,9
\label{neusup}
\end{equation}
this being the superconformal partner of the Neumann condition on $X^\mu$.
Call the ground state defined by these conditions $|0\rangle$.  In
ref.~\cite{polcai} it is shown that this corresponds to a tadpole for an RR
10-form potential (there will be more on the 10-form below).  Now go to the
mixed boundary conditions~(\ref{mixed}).  The boundary state satisfies
\begin{eqnarray}
(\psi^\mu_0 - \tilde\psi^\mu_{0}) |B\rangle &=& 0, \quad \mu=0,\ldots,p
\nonumber\\
(\psi^\mu_0 + \tilde\psi^\mu_{0}) |B\rangle &=& 0, \quad \mu=p+1,\ldots,9
\end{eqnarray}
and the ground state becomes
\begin{equation}
(\psi^{p+1}_0 + \tilde\psi^{p+1}_{0})
(\psi^{p+2}_0 + \tilde\psi^{p+2}_{0}) \ldots
(\psi^{9}_0 + \tilde\psi^{9}_{0}) |0\rangle.
\end{equation}
In the formalism of ref.~\cite{polcai} this removes $9-p$ indices, leaving a
$(p+1)$-form potential, as appropriate for coupling to a $p$-dimensional
object.  Also, since only the even forms appear in the IIb theory, and only
the odd forms in the IIa, consistency between the projections in the closed
and open string sectors (the analog of modular invariance) gives the
consistency condition stated earlier.

The actual value of the quantum of charge is of some interest.  This can be
determined from a calculation on the disk, but is more easily extracted
from a one-loop vacuum amplitude by factorization.  Consider parallel
Dirichlet $p$-branes, at $X^\mu = 0$ and at $X^\mu = Y^\mu$ for $\mu = p+1,
\ldots, 9$, where $Y^\mu$ are some fixed coordinates.  There are open strings
with one end attached to each D-brane, and the one-loop vacuum graph from
such states is a sum over cylinders with one end lying on each D-brane.
This amplitude thus also includes the exchange of a single closed string
between the two D-branes.
The amplitude is given by (we will work in Euclidean spacetime)
\begin{equation}
A = V_{p+1}\frac{1}{2} 2 \int \frac{d^{p+1}p}{(2\pi)^{p+1}}\, \sum_i
\int \frac{dt}{t}\, e^{-t(p^2 + m_i^2)/2}. \label{openloop}
\end{equation}
The factor $V_{p+1}$ is the spacetime volume of the D-brane, defined by putting
the system in a large box, the $\frac{1}{2}$ is for real fields, and the $2$ is
from interchanging the ends of the oriented string.\footnote{Alternately, the
net symmetry factor $\frac{2}{2} = 1$ arises because the discrete part of the
world-sheet diff invariance is completely fixed.}  The sum runs over the
spectrum of open strings with ends fixed on the respective D-branes; this is
given by the usual oscillator sum with an additional term $Y^\mu Y_\mu
/4\pi^2\alpha'^2$ in the mass-squared from the tension of the stretched string.
Carrying out the oscillator sum and momentum integral gives
\begin{eqnarray}
A &=& V_{p+1}  \int \frac{dt}{t}\, (2\pi t)^{-(p+1)/2}
e^{- tY^2/8\pi^2 \alpha'^2} \prod_{n=1}^\infty (1-q^{2n})^{-8} \\
&& \quad \frac{1}{2} \left\{ -16 \prod_{n=1}^\infty (1+q^{2n})^{8}
+ q^{-1} \prod_{n=1}^\infty (1+q^{2n-1})^{8}
- q^{-1} \prod_{n=1}^\infty (1-q^{2n-1})^{8}  \right\} \nonumber
\end{eqnarray}
where we define $q = e^{-t/4\alpha'}$.
The three terms in large braces come respectively from the open string R
sector with $\frac{1}{2}$ in the trace, from the NS sector with
$\frac{1}{2}$ in the trace, and the NS sector with $\frac{1}{2} (-1)^F$
in the trace; the R sector with $\frac{1}{2} (-1)^F$ gives no net
contribution.

The sum in large brackets vanishes by the usual `abstruse identity'
of supersymmetric string theory.
{}From the open string point of view this reflects the supersymmetry of the
spectrum, while in terms of the closed string exchange it reflects the fact
that there is no net force between BPS states.  As in
ref.~\cite{polcai}, it is straightforward to separate the two kinds of closed
string exchange.  Interchanging world-sheet space and time so as to see the
closed string spectrum, the terms without $(-1)^F$ in the trace come from the
closed string NS\,NS states (graviton and dilaton), while the term with
$(-1)^F$ comes from the closed string RR states.  The massless closed string
poles arise from
$t\to 0$; using standard $\vartheta$-function asymptotics in this limit, the
amplitude becomes
\begin{eqnarray}
A &=&  \frac{1}{2} (1-1) V_{p+1}  \int \frac{dt}{t} (2\pi t)^{-(p+1)/2}
(t/2\pi\alpha')^4 e^{- tY^2/8\pi^2 \alpha'^2} \nonumber\\[4pt]
&=& (1-1) V_{p+1} 2\pi (4\pi^2\alpha')^{3-p} G_{9-p}(Y^2). \label{final}
\end{eqnarray}
Here, $(1-1)$ is from the NS\,NS and RR sectors respectively, and
\begin{equation}
G_{9-p} (Y^2) = \frac{1}{4} \pi^{(p-9)/2} \Gamma\Bigl( (7-p)/2 \Bigr)
(Y^2)^{(p-7)/2}
\end{equation}
is the scalar Green function in $9-p$ dimensions.

We compare the RR contribution with that from a $(p+1)$-form potential
$A_{p+1}$,
$F_{p+2}=dA_{p+1}$, with action\footnote
{More explicitly, $F_{\mu_1 \ldots \mu_{p+2}} = (p+2) \partial_{[\mu_1}
A_{\mu_2 \ldots \mu_{p+2}]}$ and $F^*F = d^{10} x\, \sqrt{g}
F_{\mu_1 \ldots \mu_{p+2}} F^{\mu_1 \ldots \mu_{p+2}}$.}
\begin{equation}
S = \frac{\alpha_p}{2} \int F_{p+2}^*F_{p+2} + i\mu_p \int_{\rm branes}
A_{p+1}.
\label{formact}
\end{equation}
For later convenience we have not chosen a normalization for $A_{p+1}$, so
two constants $\alpha_p$ and $\mu_p$ appear.  Calculating the amplitude
from exhange of a $(p+1)$-form between the Dirichlet $p$-branes, one finds a
negative term as in the amplitude~(\ref{final}), with normalization
\begin{equation}
\frac{\mu_p^2}{\alpha_p} = 2\pi (4\pi^2\alpha')^{3-p}. \label{charge}
\end{equation}

For branes with $p + p' = 6$, the corresponding field strengths satisfy
$(p + 2) + (p' + 2) = 10$.  These are not independent in the type II string
but rather are related by Hodge duality, $F_{p+2} = {}^*F_{8 - p}$.
A Dirac
quantization condition therefore restricts the corresponding
charges~\cite{dirac}.  Integrate the field strength ${}^*F_{p+2}$ on an
$(8-p)$-sphere surrounding a
$p$-brane; from the action~(\ref{formact}) one finds total flux $\Phi=
\mu_{p}/\alpha_{p}$.
One can take ${}^*F_{p+2} = F_{8-p} = d A_{7-p}$ except on a
Dirac string at the pole.  Then
\begin{equation}
\Phi \ =\  \int_{S_{8-p}} {}^*F_{p+2} \ =\ \int_{S_{7-p}} A_{7-p}
\end{equation}
where the latter integral is on a small sphere around the Dirac string.  In
order that the Dirac string be invisible to a $(6 - p)$-brane, we need
$\mu_{6 - p} \Phi = 2\pi n$ for integer $n$.  That is, the Dirac
quantization condition is
\begin{equation}
\frac{ \mu_{p} \mu_{6 - p} }{\alpha_{p}} = 2\pi n.
\label{dirq}
\end{equation}
The charges~(\ref{charge}) of the D-branes satisfy this with minimum quantum
$n=1$.\footnote
{It follows from $F_{p+2} = {}^*F_{8-p}$ that $\alpha_{p}
= \alpha_{6 - p}$.}

{}From the point of view of the open string loop calculation this is a `string
miracle,' a coincidence in need of deeper explanation.  Had the Dirac
quantization condition not been satisfied, it would likely imply a subtle
inconsistency in the {\it type I} superstring.  That the minimum quantum is
found strongly suggests that D-branes are actually the RR-charged objects
required by string duality.

One can test this further.  While the Dirac quantization condition
constrains only the product~(\ref{dirq}), string duality makes specific
predictions for the individual charges.  Consider a $(p+1)$-dimensional
world-volume ${\cal M}$ with $p$-dimensional holes.  Under a
gauge transformation $\delta A_{p+1} = d\epsilon_{p}$, the
action~(\ref{formact}) changes by
\begin{equation}
\delta S = -i \mu_p \int_{\partial M} \epsilon_{p}.
\end{equation}
This is the change in phase of a $p$-brane state under a gauge
transformation.  In ref.~\cite{jeffandy}, the fields are normalized so that
the $2$-brane wavefunctions are invariant for $\epsilon_{2}$ being $\alpha'$
times an element of the integral cohomology.  In other words, $\mu_2 = 2\pi /
\alpha'$.  Adopting the same convention for the Dirichlet 2-branes, we would
have $\alpha_2 = 1/2\pi \alpha'^3$.  This is twice the value found in
ref.~\cite{jeffandy} (which would imply an incommensurate $\sqrt{2}$ in the
charges of the Dirichlet and solitonic 2-branes), but agrees with the
normalization in ref.~\cite{vafwit1}.\footnote{In comparing, note that
the $B_{\mu\nu}$ field in ref.~\cite{vafwit1} is twice that in
ref.~\cite{jeffandy}, with other conventions the same.}  We have not
succeeded in reconciling these calculations, but strongly expect that the RR
charge is that required by string duality.

This result for the D-brane charge is new evidence
both for string duality and for the conjecture that D-branes are the
RR-charged objects required by string duality.  That is, although it appears
that we have modified the type~II theory by adding something new to it, we
are now arguing that these objects are actually intrinsic to any
nonperturbative formulation of the type~II theory; presumably one should
think of them as an alternate representation of the black $p$-branes.  This
conjecture was made earlier and with less evidence in ref.~\cite{DLP} (the
argument there being that any object that can couple consistently to closed
string must actually be made of closed strings) and in ref.~\cite{pdinst}
(based on the
$(2n)!$ behavior of string perturbation theory~\cite{shenker}).

As an aside, this would also imply that the type~I
theory is contained within the type~II theory as a sector of the
Hilbert space.  The argument (the same as given in ref.~\cite{DLP} but now
presented in reverse order) is as follows.  Periodically identify some of the
dimensions in the type~II string,
\begin{equation}
X^\mu \sim X^\mu + 2\pi R, \quad \mu = p+1, \ldots, 9.
\end{equation}
Now make the spacetime into an orbifold by further imposing
\begin{equation}
X^\mu \sim -X^\mu, \quad \mu = p+1, \ldots, 9.
\end{equation}
To be precise, combine this with a world-sheet parity transformation to make
an orientifold~\cite{horava, DLP, vafwit2}.  This is not a consistent string
theory. The orientifold points are sources for the RR fields (by the analog of
the above arguments for D-branes, but with the boundary replaced by a
crosscap),
but in the compact space these fields have nowhere to go.  One can screen
this charge and obtain a consistent compactification with exactly 16
D-branes oriented as in eq.~(\ref{mixed}).\footnote
{The question of the consistency of orientifold compactifications has arisen.
In this paper we have encountered two necessary conditions: that the
projections in the closed string channel of the one loop open string graph
agree with the actual closed string spectrum, and that the RR forms have
consistent field equations.  We believe that these, together with the usual
modular invariance and operator product expansion closure and associativity
conditions, are also sufficient.  See refs.~\cite{lew} and in
particular~\cite{horava} for more discussion of some of these points.  This is
under further investigation~\cite{gimpol}.} Now take $R \to 0$.  The
result is the type I string~\cite{DLP,odual}.

A puzzling feature of the Dirichlet $p$-branes has always been their
diversity, with $p$ ranging from $-1$ to $9$.\footnote
{The case $p = -1$ is the D-instanton~\cite{gdinst, pdinst}.}
This now finds a
satisfying explanation in terms of the diversity of RR forms: the D-branes
comprise a complete set of electric and magnetic RR sources.  The IIa
theory has field strengths of rank $2, 4, 6, 8$ (with $n$ and $10-n$ dual),
which are the curls of potentials of rank $1, 3, 5, 7$ and so couple to
$p$-branes for $p = 0, 2, 4, 6$.  The IIb theory has field strengths of rank
$1, 3, 5, 7, 9$, which are the curls of potentials of rank
$0, 2, 4, 6, 8$ and couple to $p$-branes for $p = -1, 1, 3, 5, 7$.

The reader will notice that we have two extra branes, $p = 8$ and 9,
coupling to 9- and 10- form potentials.  While these forms do not
correspond to propagating states, they are present in the IIa and IIb
theories respectively and have important dynamical effects.  The 10-form
has been discussed previously~\cite{polcai}.  It couples to a 9-brane,
but what is that?  A 9-brane fills space, so the open string end-points are
allowed to go anywhere: this is simply a Neumann boundary condition.
If there are $n$ 9-branes (which must of course lie on top of one
another), the endpoints have a discrete quantum number: this is the
Chan-Paton degree of freedom.  The total coupling of the branes to the
10-form is
\begin{equation}
i n \mu_9 \int_{\rm spacetime} A_{10}.
\end{equation}
The equation of motion from varying $A$ implies that $n$ must equal zero.
We cannot readily cancel this with branes of the opposite orientation and
charge because we would no longer have a BPS state, but we can cancel it by
again orientifolding (with a trivial spacetime transformation) to make
the type I string.  The crosscap gives a 10-form source of the opposite sign,
giving in all
\begin{equation}
i (n - 32) \mu_9 \int_{\rm spacetime} A_{10}.
\end{equation}
Thus, the equation of motion requires the group $SO(32)$.\footnote{It is
worth recalling the logic of ref.~\cite{polcai}: the spacetime anomaly for
other groups must arise from some world-sheet superconformal anomaly, but
this must in turn correspond to some spacetime equation that is not being
satisfied.}

The 9-form potential in the IIa string has not been previously noted.  The
action $\int F_{10}{}^*F_{10}$ gives the equation of motion $d^*F_{10} = 0$,
which for a 10-form field strength implies that ${}^*F_{10}$ is constant. There
are thus no solutions at non-zero momentum, explaining why this is easily
overlooked, but the constant solution is quite interesting: it is like a
background electric field and so gives a contribution to the cosmological
constant proportional to the square of the field.\footnote{If one simply
substitutes a constant ${}^*F_{10}$ into the action one obtains a cosmological
constant of the wrong (negative) sign owing to neglect of a surface term.
It is obvious on physical grounds that the cosmological constant is positive,
and this is what one finds from the equations of motion.}
That is, the IIa superstring has a cosmological constant of undetermined
magnitude.  This is surprising, but has been partially anticipated by
Romans~\cite{romans}, who found the corresponding supergravity theory (for
fixed cosmological constant).

The implications of this are not yet clear.
Hawking~\cite{hawking} has used the same idea in four dimensions to provide
a mechanism for the variation of the cosmological constant, and then further
argued that the wavefunction of the universe forces the net low energy
cosmological constant to zero.  The latter argument hinges on aspects
of quantum gravity that are still poorly understood.

The
10-form is at first sight a violation of two pieces of string lore.
The first as that there are no free parameters in string theory:
the value of $F_{10}$ is midway between a field and a parameter, being
spacetime-independent but evidently determined by initial conditions.\footnote
{Nucleation of 9-branes shifts the 10-form field strength by a large
discrete unit, by analogy
with two-dimensional massive electrodynamics.}
The second is that it is not possible to break
supersymmetry at tree level with a continuous parameter: the supersymmetry
transformations contain terms of order $F_{10}$ ($m$ in the notation of
ref.~\cite{romans}) which make at least some previously supersymmetric
states nonsupersymmetric.  However, it is possible that the value of $F_{10}$
will turn out to be quantized in string units, at least in some
compactifications.
This question, and the question of how this and other RR backgrounds affect
physics in four dimensions, are very interesting and are under
investigation~\cite{joeandy}.

\subsection*{Acknowledgments}

I would like to thank Michael Green and Edward Witten for useful discussions,
Constantin Bachas and Shanta de Alwis for corrections to the manuscript,
and Andy Strominger for many discussions and for bringing ref.~\cite{romans} to
my attention.  This work is supported by NSF grants PHY91-16964 and
PHY94-07194.


\begin{thebibliography}{99}

\bibitem{GSW}
M. B. Green, J.H. Schwarz, E. Witten, {\it Superstring Theory,}
Cambridge, UK (1987).

\bibitem{FMS}
D. Friedan, E. Martinec, and S. Shenker, Nucl. Phys. {\bf B271}, 93 (1986).

\bibitem{RRch}
C. M. Hull and P.K. Townsend,
Nucl. Phys. {\bf B438}, 109 (1995);\\
E. Witten, Nucl. Phys. {\bf B443}, 85 (1995);\\
A. Strominger, {\it Massless Black Holes and Conifolds in String Theory,}
hep-th/9504090 (1995).

\bibitem{black}
G. T. Horowitz and A. Strominger,
Nucl. Phys. {\bf B360}, 197 (1991).

\bibitem{DLP}
J. Dai, R. G. Leigh, and J. Polchinski,
Mod. Phys. Lett. {\bf A4}, 2073 (1989);\\
R. G. Leigh, Mod. Phys. Lett. {\bf A4}, 2767 (1989).

\bibitem{odual}
P. Horava, Phys. Lett. {\bf B231}, 251 (1989);\\
M. B. Green, Phys. Lett. {\bf B266}, 325 (1991).

\bibitem{gdinst}
M. B. Green, Phys. Lett. {\bf B329}, 435 (1994).

\bibitem{pdinst}
J. Polchinski, Phys. Rev. {\bf D50}, 6041 (1994);\\
M. B. Green, Phys. Lett. {\bf B354}, 271 (1995);\\
M. B. Green, {\it Boundary Effects in String Theory,}
preprint hep-th/9510016 (1995).


\bibitem{callan}
C. Lovelace, Phys. Lett. {\bf B34}, 500 (1971);\\
L. Clavelli and J. Shapiro, Nucl. Phys. {\bf B57}, 490 (1973);\\
M. Ademollo, R. D' Auria, F. Gliozzi, E. Napolitano, S. Sciuto, and P. di
Vecchia, Nucl. Phys. {\bf B94}, 221 (1975);\\
C. G. Callan, C. Lovelace, C. R. Nappi, and S. A. Yost,
Nucl. Phys. {\bf B293}, 83 (1987).

\bibitem{polcai}
J. Polchinski and Y. Cai, Nucl. Phys. {\bf B296}, 91 (1988);\\
C. G. Callan, C. Lovelace, C. R. Nappi and S.A. Yost,
Nucl. Phys.  {\bf B308}, 221 (1988).

\bibitem{horava}A. Sagnotti,  {\it Non-Perturbative Quantum
Field Theory,}
eds. G. Mack et. al. (Pergamon Press, 1988), p. 521;\\
M. Bianchi and A. Sagnotti, Phys. Lett. {\bf 247B}
(1990) 517; Nucl. Phys. {\bf B361} (1991) 519;\\
P. Horava, Nucl. Phys. {\bf B327}, 461 (1989).

\bibitem{dirac}
R. I. Nepomechie, Phys. Rev. {\bf D31}, 1921 (1985);\\
C. Teitelboim, Phys. Lett. {\bf B167}, 63, 69 (1986).

\bibitem{jeffandy}
J. A. Harvey and A. Strominger, Nucl. Phys. {\bf B449}, 535 (1995).

\bibitem{vafwit1}
C. Vafa and E. Witten, Nucl. Phys. {\bf B447}, 261 (1995).

\bibitem{shenker}
S. H. Shenker, in {\it Cargese 1990, Proceedings: Random Surfaces and Quantum
Gravity,} 191 (1990).

\bibitem{vafwit2}
C. Vafa and E. Witten, {\it Dual String Pairs with $N=1$ and $N=2$
Supersymmetry in Four Dimensions,} HUTP-95-A023, hep-th/9507050 (1995).

\bibitem{lew}
D. C. Lewellen, Nucl. Phys. {\bf B372}, 654 (1992).

\bibitem{gimpol}
E. Gimon and J. Polchinski, work in progress.

\bibitem{romans}
L. J. Romans, Phys. Lett. {\bf B169}, 374 (1986).

\bibitem{hawking}
S. W. Hawking, Phys. Lett. {\bf B134}, 403 (1984).

\bibitem{joeandy}
J. Polchinski and A. Strominger, work in progress.

\end{thebibliography}
\end{document}